\documentclass[numberedheadings]{aipproc}
\layoutstyle{6x9}
\usepackage{epsfig,rotating}

\def\nubar{$\overline{\nu}\ $}
\def\nue{\ensuremath{\nu_{e}}}
\def\nubare{\ensuremath{\overline{\nu}_{e}}}

\def\numu{\ensuremath{\nu_{\mu}}}
\def\nubarmu{\ensuremath{\overline{\nu}_{\mu}}}

\newcommand{\pnuenumu}{\ensuremath{p(\nue \rightarrow \numu)\,}}
\newcommand{\nuenumu}{\ensuremath{\nue \rightarrow \numu\,}}
\newcommand{\nubarenubarmu}{\ensuremath{\overline{\nu}_e \rightarrow \overline{\nu}_\mu\,}}
\newcommand{\dmot}{\ensuremath{\delta m^2_{12}\,}}
\newcommand{\dmtt}{\ensuremath{\delta m^2_{23} \,}}

\newcommand{\He}{\ensuremath{^6{\mathrm{He}\,}}}
\newcommand{\Ne}{\ensuremath{^{18}{\mathrm{Ne}\,}}}

\newcommand{\thetaot}{\ensuremath{\theta_{13}}\,}

\newcommand{\numunue}{\ensuremath{\nu_\mu \rightarrow \nu_e}}

\newcommand{\sigdm}{\ensuremath{{\rm sign}(\Delta m^2)}}

\begin{document}
\title{Beta-Beams: present design and expected performances\footnote{
Talk presennted at
NuFact 03, 5$^{\mathrm{th}}$ International Workshop on Neutrino Factories \& Superbeams,
5--11 June 2003, Columbia University, New York.}}

\author{Jacques Bouchez}{address={DAPNIA, CEA Saclay, France}}
\author{Mats Lindroos}{address={CERN, Geneva, Switzerland}}
\author{Mauro Mezzetto}
{address={Istituto Nazionale Fisica Nucleare, Sezione di Padova, Italy.},
email={mezzetto@pd.infn.it}}
\begin{abstract} 
We give the present status of the beta-beam study, which aims at producing
intense \nue\  and \nubare\  beams from the decay of relativistic radioactive ions.
The emphasis is put on recent technical progress and new ideas. The expected
performances in terms of neutrino mixing parameters ($\theta_{13}$ and CP
violating phase $\delta$) using a megaton water Cerenkov detector installed
in the Fr{\'e}jus underground laboratory are shown to be excellent, and the
synergy with a a companion SuperBeam is underlined. 
\end{abstract}

\maketitle

\section{Motivations}

Super-Kamiokande has given strong evidence for a maximal oscillation between
\numu and $\nu_\tau$ \cite{SKatm}, and several projects with accelerators have
been designed to check this result. The first results of the K2K
experiment \cite{K2K} confirm the oscillation, and future projects (MINOS 
in the USA, OPERA and ICARUS at Gran Sasso) should refine the oscillation 
parameters by 2010.

More recently, after the results from SNO \cite{SNO-salt}
and Kamland \cite{KamL}, a solid
proof for solar neutrino flavour oscillations governed by the
so-called LMA solution has been established. We can no longer
escape the fact that neutrinos have indeed a mass, although the
absolute scale is not yet known. Furthermore, the large mixing
angles of the two above-mentioned oscillations and their relative
frequencies open the possibility to test CP violation in the
neutrino sector if the third mixing angle, $\theta_{13}$, is not
vanishingly small (we presently have only an upper limit set at
10 degrees on $\theta_{13}$, provided by the 
CHOOZ experiment \cite{CHOOZ}). Such a
violation could have far reaching consequences, since it is
a crucial ingredient of leptogenesis, one of the
presently preferred explanations for the matter dominance in our
Universe.

The ideal tool for these studies is thought to be the so-called neutrino
factory, which would produce through muon decay intense neutrino beams
aimed at magnetic detectors placed several thousand kilometers away from
the neutrino source.

 However, such projects would not be launched unless
one is sure that the mixing angle $\theta_{13}$, governing the oscillation
between \numu and \nue at the higher frequency, is such that this oscillation
is indeed observable.
This is why physicists have considered the possibility
of producing new conventional neutrino beams of unprecedented intensity, made
possible by recent progress on the conception of proton drivers with a factor
10 increase in power (4 MW compared to the present 0.4 MW of the FNAL beam)
The present limit on $\theta_{13}$ is 10 degrees, these new neutrino
``superbeams'' would explore $\theta_{13}$ down to 1 degree (i.e a factor
100 improvement on the \numu - \nue  oscillation amplitude).

European working groups have studied a neutrino factory at CERN
for some years, based on a new proton driver of 4 MW, the SPL.
Along the lines described above, a subgroup on neutrino
oscillations has studied the potentialities of a neutrino
SuperBeam produced by the SPL. The energy of produced neutrinos is
around 270 MeV, so that the ideal distance to study \numu to \nue
oscillations happens to be 130 km, that is exactly the distance
between CERN and the existing Fr{\'e}jus laboratory. The present
laboratory cannot house a detector of the size needed to study
neutrino oscillations, which is around 1 million cubic meters. But
the recent decision to dig a second gallery, parallel to the
present tunnel, offers a unique opportunity to complete the needed
extension in 2012 for a reasonable price.

Due to the schedule of the new gallery, a European project would
be competitive only if the detector at Fr{\'e}jus reaches a
sensitivity on $\theta_{13}$ around 1 degree, since other projects in
Japan (JHF phase 1) and USA (NuMI off-axis) will have reached 2.5
degrees by 2013. The working group has then decided to study
directly a water \v{C}erenkov detector with a mass around 1 megaton,
necessary to reach the needed sensitivity. It has benefited from a
similar study by our American colleagues, the so-called UNO
detector \cite{UNO} with a total mass of 660 kilotons. Simulations have shown
that the sensitivity on $\theta_{13}$ at a level of 1 degree could indeed
be fulfilled. However, the study of CP violation requires the SPL to be
run sequentially with neutrinos and antineutrinos, and due to the fact
that less antineutrinos are produced (less $\pi^-$ than $\pi^+$ are produced)
and that the \nubar cross section is 5 times lower at the considered energies,
10 years of running should be shared roughly in 2 years with $\nu$ and 8 years
with \nubar. This would be a strong limitation on CP sensitivity.

 This is where the beta beam concept, initially proposed by 
Piero Zucchelli \cite{Piero},
comes into play. The idea is to produce well collimated and intense
\nue (\nubare) beams by producing, collecting, accelerating to energies with
$\gamma$ factor around 100 and storing in a final decay ring radioactive ions
chosen for their ability to be copiously produced and with a lifetime around 1
second. The best candidates happen to be $^{18}Ne$ for \nue and $^6He$ for
\nubare. A baseline study for such a BetaBeam complex has been produced at
CERN \cite{Lindroos}, where there is a strong expertise on ion beams, 
both for nuclear physics
through ISOLDE and for high energy experiments. 

The initial goal was to produce a \nue beam which could be run simultaneously
with the \numu SPL SuperBeam, so that 10 years of data could be accumulated
for the each of the 2 time reversed oscillations, \nuenumu and \numunue .
This project was already presented at NuFact02 \cite{SPL-SB,myoldbeta}.

A workshop took place in march 2003 at Les Arcs \cite{betawork}, where 
nuclear physicists (mainly those concerned with the EURISOL project), 
neutrino physicists and machine scientists have met to discuss BetaBeam issues.
The aim of this workshop was to get updated on recent progress on the
BetaBeam project, explore the synergies between beta beams and
EURISOL~\cite{Eurisol}, and
identify common studies which could benefit to both communities.

Apart from new ideas which simplify the overall BetaBeam design, the major
innovation was the proposal to run simultaneously with both types of ions
($\beta^+$ and $\beta^-$ emitters) stored in the ring. This opens up the
exciting possibility of performing efficiently CP violation studies with
beta beams alone, and get very useful redundancies by comparing SuperBeam and
BetaBeam data.

The section 2 describes the machine aspects of beta beams,
the section 3 gives the expected performances on the measurement of 
the mixing angle $\theta_{13}$ and the CP violating $\delta$ phase.

\section{The BetaBeam complex}
\begin{figure}[ht]
\centerline{\epsfig{file=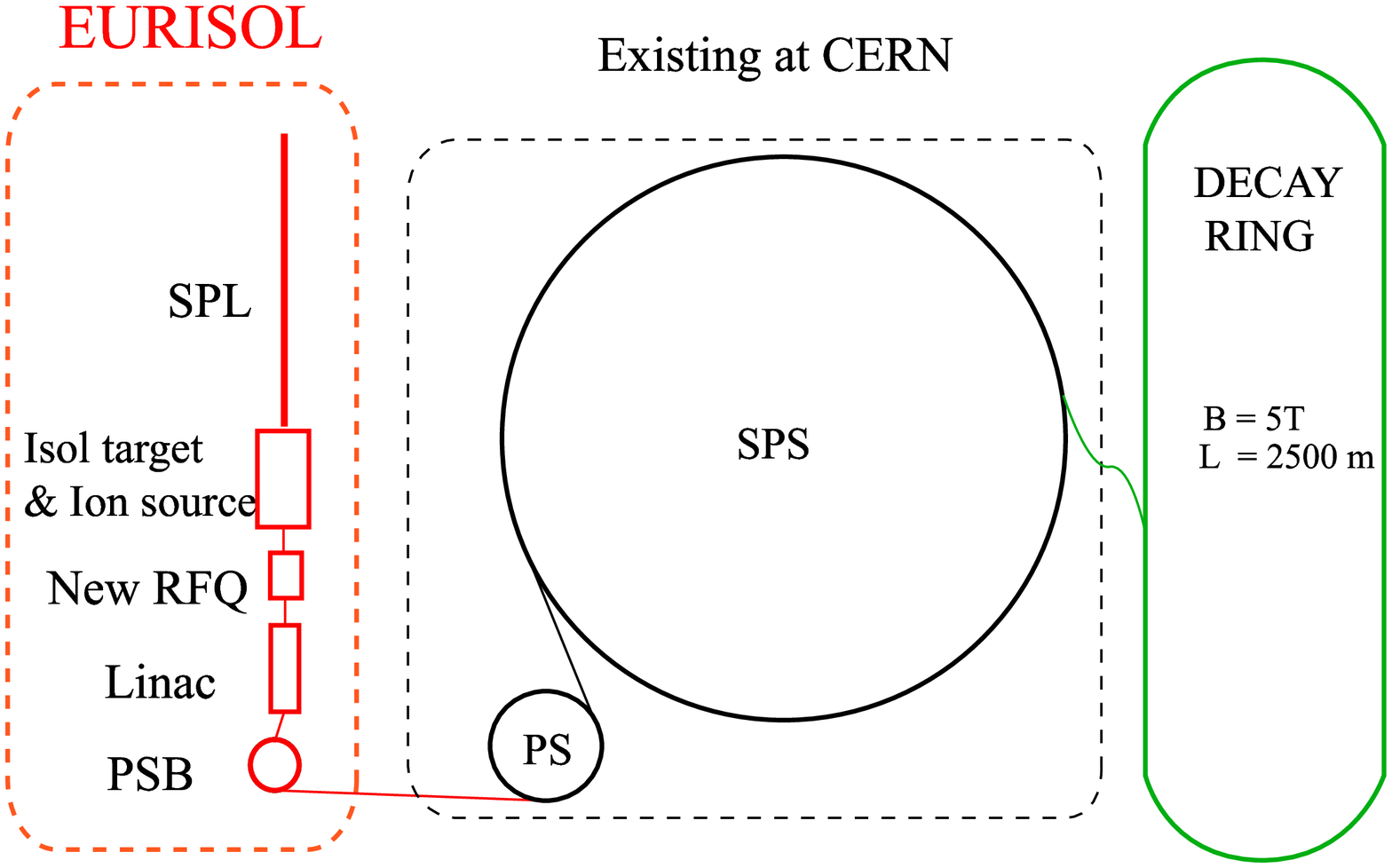,width=0.70\textwidth}  }
\caption{Schematic layout of the beta-beam complex. At left, the low energy part is
largely similar to the EURISOL project. The central part (PS and SPS) uses
existing facilities. At right, the decay ring has to be built.}
\label{fig:sketch}
\end{figure}

The beta-beam complex is shown schematically on figure~\ref{fig:sketch}.
Technical details and recent progress on this project can be found at
{\tt http://beta-beam.web.cern.ch/beta-beam/}

 The protons are delivered by the Super Proton Linac (SPL) \cite{SPL},
 which is being 
studied at CERN in the framework of the neutrino factory \cite{Nufact}. Such an intense
proton driver would deliver 2mA of 2.2 GeV (kinetic energy) protons hopefully
by 2012.
 An ISOL target would need only 100 $\mu$A, that is 5 \% only of the total
proton intensity.
\subsection{The target and the ion source}
 The targets are similar to the ones envisioned by EURISOL~\cite{Eurisol}:
 for $^6$He, it consists 
either of a water cooled  tungsten core or of
a liquid lead core which works as a proton to neutron converter
surrounded by beryllium oxide \cite{Nolen} ,aiming for 10$^{15}$
fissions per second. 
$^{18}$Ne can be produced by spallation reactions,
in this case protons will directly hit a magnesium 
oxide target.
The collection and ionization of the ions is performed using the ECR technique.
The pulsed "ECR-duoplasmatron" under development at Grenoble, using very dense plasmas
( 10 $^{14}$/cm$^3$) and high magnetic fields (2 to 3 Teslas) submitted to 
high frequencies (60 to 90 GHz) is aimed to produce 10$^{12}$ to
10$^{13}$ ions in very short bunches (20 to 100 $\mu$s) at 100 keV 
with repetition rates
reaching 16 Hz \cite{Sortais}. 
The advantage over the standard ECR technique is that the
downstream complex can be simplified, due to the achieved bunching and
hopefully to ions which are totally ionized. 
\subsection{First acceleration and storing}
Then the first acceleration process can be achieved using a LINAC rather than a
cyclotron or a FFAG as initially considered. Ions would be accelerated to 20-100
MeV/u in 16 batches per second. Then comes the first storage ring, a rapid
cycling synchrotron using multiturn injection (40 turns), delivering a single
150 ns bunch at 300 MeV/u.
\subsection{Final acceleration}
16 bunches (consisting of 2.5 10$^{12}$ ions each in the case of He) are then
accumulated into the PS, and reduced to 8 bunches during their acceleration
to intermediate energies.
Due to the fact that the PS is a slow machine, this is the
place where radiation levels due to ion decays is the most severe (the
replacement of the PS by a more rapid machine would ease this problem and
many others in the CERN complex of accelerators). Furthermore, the space charge
bottleneck at SPS injection will require a transverse emittance blow-up.
The SPS will finally accelerate the 8 bunches to the desired energy
($\gamma \simeq$ 100) using a new 40 MHz RF system and the existing 200 MHz RF
system, before ejecting them in batches of four 10 ns bunches 
into the decay ring.
\subsection{The decay ring}
\label{sec:Decay Ring}
This ring has the shape of an hippodrome, with a total length of 6880 m
(matching the SPS) and straight sections of 2500 m each (36\%).
Due to the relativistic time dilatation, the ion lifetimes reach several
minutes, so that stacking the ions in the decay ring is mandatory to get enough
decays and hence high neutrino fluxes. The challenge is then to inject ions
in the decay ring and merge them with existing high density bunches.
As conventional techniques with fast elements are excluded, a new scheme 
(asymmetric merging) was specifically conceived for this task
\cite{demo}. 
It schematically consists in injecting an off-momentum ion bunch 
on a matched dispersion
trajectory, then rotate this fresh bunch in longitudinal phase space by a
1/4 turn into a starting configuration for bunch merging. This technique had
been proven to work on computer simulations \cite{demo}, 
but it received very recently
a first experimental confirmation \cite{mergexp}.
\subsection{Neutrino fluxes}
\label{sec:fluxes}
One of the ideas presented at the Moriond meeting was that it should be 
possible to run together Neon and Helium ions in the decay ring (of course,
in different bunches). Due to their different rigidities, these ions would
have relativistic $\gamma$ factors in the 5 to 3 ratio, which is quite
acceptable for the physics program. This will impose constraints on the 
lattice design for the decay ring, but no impossibility has been identified.

An ECR source coupled to an EURISOL target would 
produce $2 \cdot 10^{13}$ $^6$He ions per second.
Taking into account all decay losses along the accelerator complex, and
estimating an overall transfer efficiency of 50\%, one estimates that
$4 \cdot  10^{13}$ ions would permanently reside in the final decay ring for
$\gamma$ = 60. That would give an antineutrino flux aimed at the Fr{\'e}jus
underground laboratory of $2.1 \cdot 10^{18}$ per standard year (10$^7$ s).

For $^{18}$Ne, the yield is expected to be only $8\cdot 10^{11}$ ions per second.
Due to this smaller yield, which could be certainly improved with some R\&D,
it was then proposed to use 3 EURISOL targets in sequence connected to the
same ECR source.
Again taking into account decay losses plus a 50\% efficiency, this means that
$2\cdot 10^{13}$ such ions would reside in the decay ring for $\gamma$ = 100, 
giving rise to a neutrino flux of $0.35 \cdot 10^{18}$ per standard year.  

All these numbers are preliminary and need to be refined. They are however
based on the present state of the art for the technology, and 
suppose using the present PS, while the SPS cycle is set at 16 s;
a shorter cycle for the SPS would improve the 
accumulation factor substantially, while a faster PS would increase the
intensity of ions making it to the decay ring.

In the following study, it was supposed that the neutrino flux from \Ne 
could be increased
by a factor 3 over the present conservative estimate, having
room for improvements both
in the cycle duration of PS and SPS and in
the \Ne production at the targets with a dedicated R\&D, while only a 40 \% 
improvement was put on antineutrino fluxes. 

\subsection{Radiation issues}
The main losses are due to decays of He ions, and reach 1.2 W/m in the PS
and 9 W/m in the decay ring. This seems manageable, although the use of
superconducting bending magnets in the decay ring requires further studies.
Activation issues have been recently addressed \cite{Magi}, and show that
the dose rate on magnets in the arcs is limited to 2.5 mSv/h at contact after
30 days operation and 1 day cooling. Furthermore, the induced radioactivity
on ground water will have no impact on public safety.

\section{Physics reach}
The following study is based on the hypothesis that a UNO-like water Cerenkov
detector (440 kt fiducial mass) will be installed in the underground
Fr{\'e}jus laboratory and receive neutrino beams produced at CERN, 130 km away.
\subsection{Signal and backgrounds} 
\begin{figure}[ht]
    \begin{minipage}{\textwidth}
      \begin{minipage}{0.56\textwidth}
      \centerline{\epsfig{file=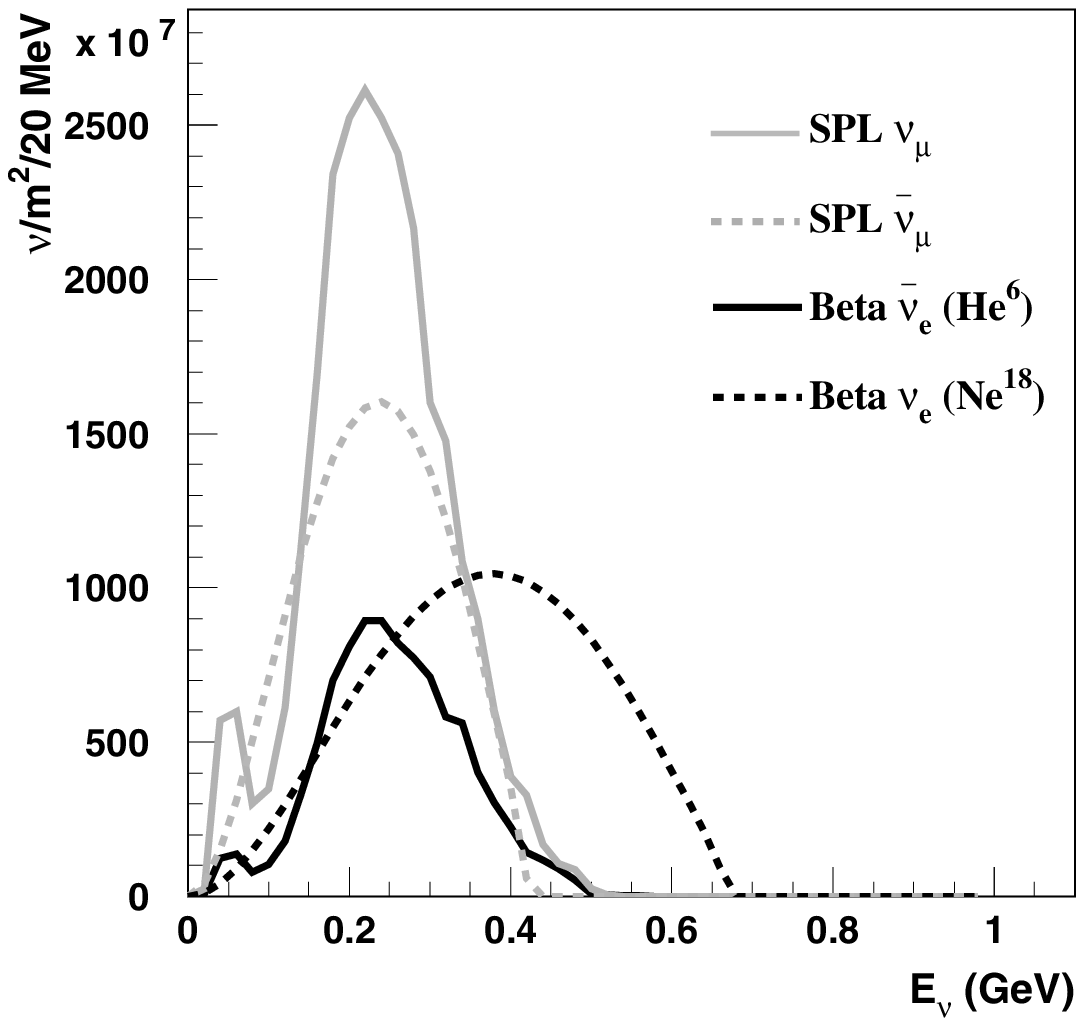,width=0.9\textwidth}  }
      \end{minipage}
      \begin{minipage}{0.43\textwidth}
      \begin{tabular}{ccc}
      \hline
           &   Fluxes       & $\langle E_\nu \rangle$  \\
           &  $\nu/m^2/yr$  & (GeV)  \\
      \hline
          \nubare ($\gamma=60$)  &   $1.97 \cdot 10^{11}$ &  0.24 \\ 
          \nue ($\gamma=100$)    & $1.88 \cdot 10^{11}$ &  0.36  \\
      \hline
          \numu    &     $4.78 \cdot 10^{11}$ &  0.27  \\ 
          \nubarmu &     $3.33 \cdot 10^{11}$ &  0.25  \\ 
      \end{tabular}
      \end{minipage}
    \end{minipage}
    \caption{Beta Beam fluxes at the Frejus location  (130 km baseline).
     Also the SPL Super Beam \numu \  and \nubarmu \  fluxes are shown in the
    plot.}
    \label{fig:fluxes}
\end{figure}

The neutrino beam energy depends on
the $\gamma$ of the parent ions in the decay ring. 
As discussed in ref.~\cite{myoldbeta}, the optimization of this 
energy, is a compromise between the advantages of the higher $\gamma$, as a
 better focusing, higher cross sections and higher signal efficiency;
 and the advantages of the lower $\gamma$ 
values as the reduced background rates (see the following)
 and the better match with the
probability functions. 
Given the decay ring  constraint (see sect.~\ref{sec:Decay Ring}):
$\gamma(^6{\rm He})/\gamma(^{18}{\rm Ne})=3/5$ 
the optimal $\gamma$  values result to be $\gamma(\He)=60$ and
$\gamma(\Ne)=100$.
A flux of $2.9 \cdot 10^{18}$ \He\ 
decays/year and $1.1\cdot10^{18}$ \Ne\  decays/year, as
discussed in sect.~\ref{sec:fluxes}, will be assumed.
Fig.~\ref{fig:fluxes} shows the BetaBeam neutrino fluxes computed at the
130 Km baseline, together with the SPL Super Beam (SPL-SB).

The mean neutrino energies of the \nubare, \nue\  beams are 0.24 GeV
 and 0.36 GeV respectively.
 They are well matched with the CERN-Frejus 130 km baseline.
On the other hand energy resolution is very poor at these energies, given
the influence of Fermi motion and other nuclear effects and in the following
all the sensitivities are computed for a counting experiment with no
energy cuts.

The signal in a Beta Beam looking for \nuenumu oscillations would be the
appearance of \numu\  charged-current events, mainly via quasi-elastic
interactions. These events are selected by requiring
        a single-ring event,
        the track identified as a muon using the standard 
Super-Kamiokande identification algorithms (tightening the
cut on the pid likelihood value), and
        the detection of the muon decay into an electron.
Background rates and signal efficiency have been studied in a
full simulation, using the NUANCE code~\cite{casper}, reconstructing
events in a Super-Kamiokande-like detector. 

The Beta Beam is intrinsically free from contamination by any different
flavor of neutrino. However, background can be generated by inefficiencies
in particle identification, such as mis-identification of pions produced
in neutral current single-pion resonant interactions,
electrons (positrons) mis-identified as muons, or by external sources such as
atmospheric neutrino interactions.

The pion background has a threshold at neutrino energies of about 450 MeV,
and is highly suppressed at the Beta Beam energies. The electron background is
almost completely suppressed by the request of the detection of a delayed
Michel electron following the muon track.
 The atmospheric neutrino
background can be reduced mainly by timing the parent ion bunches.
 For a decay ring straight sections of
2.5~km and a bunch length of 10~ns, which seems feasible \cite{Lindroos}, 
this background becomes negligible \cite{myoldbeta}.
Moreover, out-of-spill neutrino interactions can be used to
normalize this background to the 1\% accuracy level.

Signal and background rates for a 
4400~kt-yr exposure to  $^6$He and $^{18}$Ne beams, together with the
SPL SuperBeam (SPL-SB) fluxes \cite{SPL-SB}, are reported in table~\ref{tab:beta:rates} 
\begin{table}
\caption{\label{tab:beta:rates}
 Event rates for a 4400~kt-y exposure. The signals are
computed for \mbox{$\thetaot=3^\circ$}, 
$\delta=90^\circ$ $\sigdm=+1$. ``$\delta$-oscillated'' events indicates the difference between
the oscillated events computed with $\delta=90^\circ$ and with
$\delta=0$. ``Oscillated at the Chooz limit'' events are computed for
$\sin^2{2\theta_{13}}=0.12$, $\delta=0$.}
\begin{tabular}{@{}lrrrr}
                & \multicolumn{2}{c}{Beta Beam} & \multicolumn{2}{c}{SPL-SB}\\
 		& $^6He$($\gamma=60$)  &   $^{18}Ne$($\gamma=100$) & \numu (2 yrs) & \nubarmu (8 yrs) \\
\hline
CC events (no osc, no cut) & 19710   & 144784 & 36698  & 23320\\
Oscillated at the Chooz limit & 612 & 5130   & 1279 & 774  \\
Total oscillated ($\delta=90^\circ$, $\thetaot=3^\circ$)&  44     &  529   & 93     & 82\\
$\delta$ oscillated          & -9      &  57    & -20    & 12\\
Beam background            &  0      &  0     & 140    & 101\\
Detector backgrounds       &   1     &  397   &  37    & 50\\
\end{tabular}
\end{table}
\subsection{Systematic errors}

A facility where the neutrino fluxes are known with great precision
is the ideal place where to measure neutrino cross sections.
In the Beta Beam the neutrino fluxes are completely defined by the
parent ions beta decay properties and by the number 
of ions in the decay ring.
A close detector of $\sim 1~$kton placed at a distance of 
about 1~km from the decay ring could then measure the relevant
neutrino cross sections. 
Furthermore the $\gamma$ factor of the accelerated ions can be varied.
 In particular
a scan can be initiated below the background production threshold,
allowing a precise measurement of the cross sections for resonant processes.
It is estimated that a residual systematic error of 2\% will be
the final precision with which both the signal and the backgrounds
can be evaluated.

The \thetaot and $\delta$ sensitivities are computed taking into account
        a 10\% error on the solar $\delta m^2$ and $\sin^2{2\theta}$,
already reached after the recent SNO-salt results \cite{SNO-salt}
  and a 5\% and 1\% error on $\delta m^2_{23}$ and $\sin^2{2\theta_{23}}$
respectively,
as expected from the J-Parc neutrino experiment~\cite{jhf}.
Only the diagonal contributions of these errors are considered.
In the following the default values for the oscillation parameters
will be \mbox{$\sin^2{2\theta_{23}}=1$}, \mbox{$\dmtt=2.5\cdot 10^{-3} {\rm eV}^2$},
 \mbox{$\sin^2{2\theta_{12}}=0.8$}, \mbox{$\dmot=7.1\cdot 10^{-5} {\rm eV}^2$},
 \sigdm=+1.

\subsection{Parameter correlations and degeneracies}
\begin{figure}[ht]
\centerline{\epsfig{file=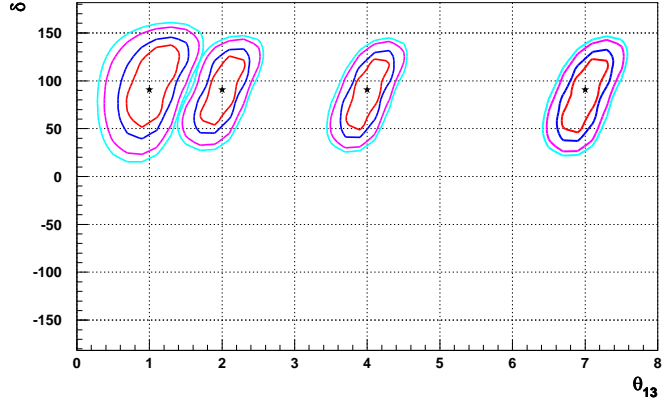,width=0.60\textwidth}  }
\caption{Fits to \thetaot and $\delta$ after a 10 yrs 
BetaBeam run. Plots are shown for $\delta=1^\circ,2^\circ,4^\circ,7^\circ$.
For the other neutrino oscillation parameters see the text.
Lines show $1 \sigma$, 90\%, 99\% and $3 \sigma$ confidence levels.  }
\label{fig:many_plots}
\end{figure}

Correlations between \thetaot and $\delta$ are fully accounted for,
and indeed they are negligible as can be
seen in the fits to \thetaot and $\delta$ shown in Fig.~\ref{fig:many_plots}.

The net effect of the \sigdm ambiguity is to make undetectable
sign($\delta \cdot \sigdm$). This derives by the negligible
matter effects at the 130 km, so that a change of \sigdm is
equivalent to a change of the sign of $\delta$.
The performances of the BetaBeam to the two opposite values of
sign($\delta \cdot \sigdm$) are different because the neutrino and
antineutrino runs have different statistics and backgrounds.
This effect will be illustrated in fig.~\ref{fig:beta:sens} and 
fig.~\ref{fig:CP:delta}.

Finally the $\theta_{23}/(\pi/2-\theta_{23})$ ambiguity 
is formally taken into account, but no effect is found because
 the BetaBeam performances
are computed for the central value of SuperKamiokande:
$\theta_{23}=45^\circ$.
A study of the performances of
the BetaBeam for different values of $\theta_{23}$ is beyond
the purpose of this article.

We stress the fact that an experiment working at very short baselines
has the smallest possible parameter degeneracies and 
ambiguities and it is the cleanest possible environment where to look
for genuine leptonic CP violation effects. 

\subsection{$\thetaot/\delta$ sensitivities}

\begin{figure}[ht]
        {\epsfig{file=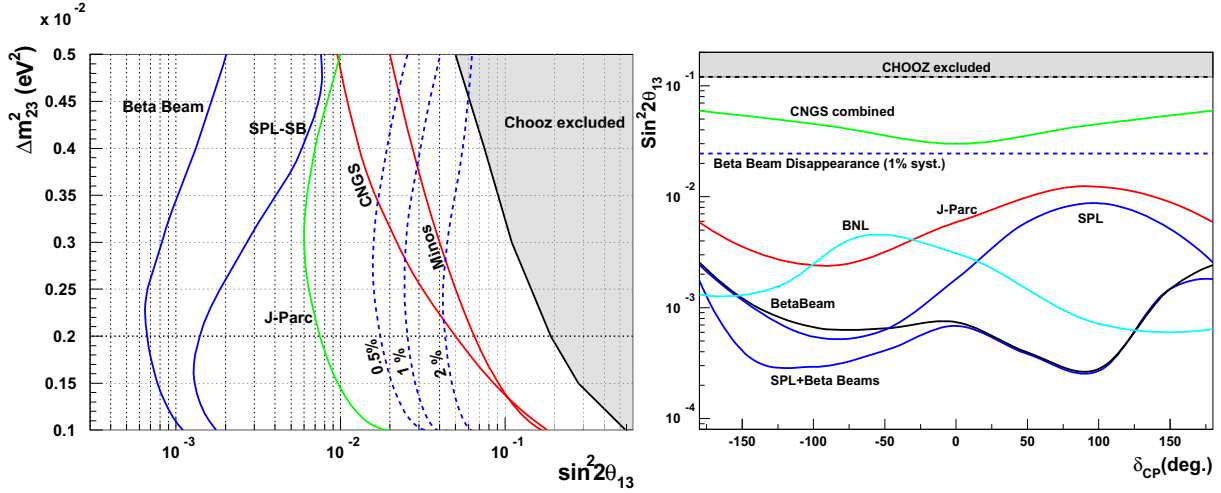,width=1.1\textwidth}  }
  \caption{LEFT: 90\%CL sensitivity of the disappearance channel to \thetaot
  in a 5 yrs run drawn as dotted lines. The
  labels 0.5\%, 1\% and 2\% indicate the systematic errors with
  which are computed. Also shown are the appearance sensitivities of Beta
  and SPL beams, computed for $\delta=0$, \sigdm=+1. The combined
  CNGS limit is taken from ref.~\cite{Migliozzi}, J-Parc from~\cite{jhf},
  Minos from ref.~\cite{Tzanakos}.
  RIGHT: 90\%CL sensitivity expressed as function of $\delta$ for
  $\dmtt=2.5\cdot10^{-3}eV^2$.
  CNGS and J-Parc curves are taken from ref.~\cite{Migliozzi}, BNL from
  ref.~\cite{Diwan}. All the appearance sensitivities are computed for
  $\sigdm=+1$.}
  \label{fig:th13}
\end{figure}

The \thetaot angle can be independently explored both with \nue and \nubare
disappearance measurements.
We note that the comparison of the \nue and \nubare disappearance
experiments could set limits to CPT violation effects.
 Sensitivities to \thetaot, computed for a 5 yr run and
for systematic errors equal to 2\%, 1\% and 0.5\% are shown 
Fig.~\ref{fig:th13}left).
For comparison sake, shown in the same plot are the sensitivities
reachable with the appearance channels, computed for $\delta=0$.

Indeed \thetaot and $\delta$ are so tightly coupled in the appearance
channels 
that the sensitivity expressed for $\delta=0$ is purely
indicative. A better understanding of the sensitivity of the BetaBeam is
expressed in the $(\thetaot,\delta)$ plane, having fixed all the
other parameters ($\dmtt=2.5 \cdot 10^{-3}$ eV$^2$), as shown in 
Fig.~\ref{fig:th13}right).
In the same plot the sensitivity of the SPL-SB computed
for a 5 yrs \numu run is displayed . It can be noted the very large variation
of the SPL-SB sensitivity for the different values of $\delta$,
characteristic of the single flavour run. The BetaBeam, having both
CP neutrino states in the same run, exhibits a much more favourable
dependence to the CP phase $\delta$.

A search for leptonic CP violation can be performed running the Beta Beam
with $^{18}$Ne and $^6$He, and fitting the number of muon-like events to
the \pnuenumu and to the p(\nubarenubarmu) probabilities.
 The fit can provide the simultaneous determination
of $\theta_{13}$ and $\delta$, see fig.~\ref{fig:many_plots}.

\begin{figure}[ht]
\centerline{\epsfig{file=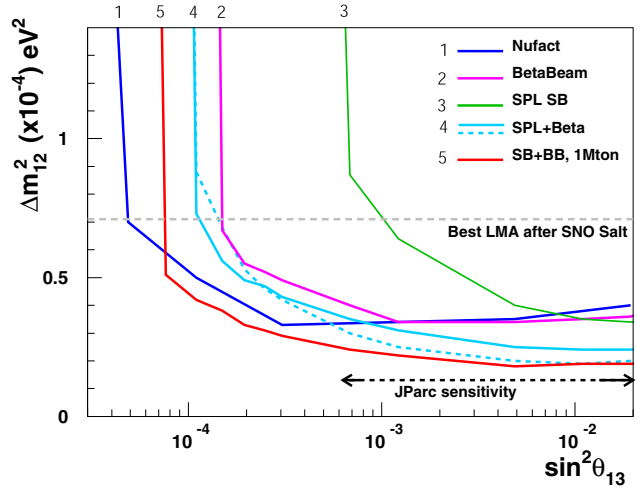,width=0.60\textwidth}  }
\caption{99\%CL $\delta$ sensitivity of the Beta Beam, of the SPL-SuperBeam, 
and of their combination, see text. 
Dotted line is the combined SPL+Beta sensitivity computed for
\sigdm=-1.
 Sensitivities are 
 compared with a 50 GeV Neutrino Factory 
producing       $2\cdot10^{20} \mu$ decays/straight section/year, 
and two 40 kton detectors at 3000 and 7000 km \cite{golden2}.  }
\label{fig:beta:sens}
\end{figure}

Event rates are summarized in Table~\ref{tab:beta:rates}.
The region of 99\% CL  sensitivity to maximal CP violation ($\delta=90^\circ$)
in the \dmot and \thetaot parameter space, following the convention of~\cite{golden2},
is plotted in Fig.~\ref{fig:beta:sens}.

\begin{figure}[th]
\centerline{\epsfig{file=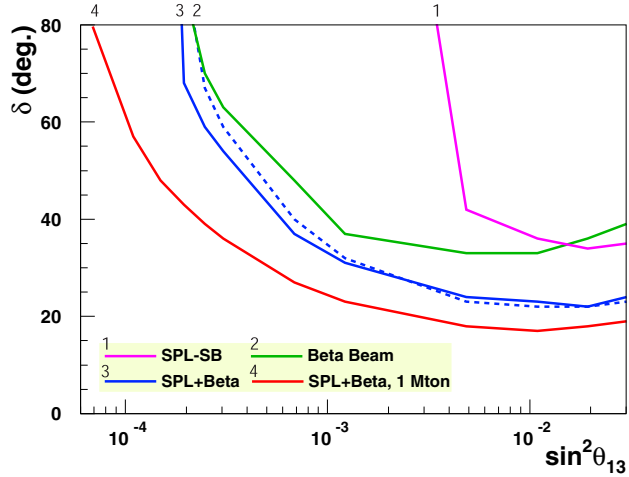,width=0.60\textwidth} }
\caption{$\delta$ discovery potential  ($3 \sigma$) 
as function of \thetaot. Dotted line are sensitivities
computed for \sigdm=-1}
\label{fig:CP:delta}
\end{figure}

The $3 \sigma$ sensitivity to $\delta$, having fixed 
$\dmot=7.1\cdot10^{-5}$ eV$^2$, is shown in Fig.~\ref{fig:CP:delta}.

\subsection{Synergies between the SPL-SuperBeam and the Beta Beam}

The Beta Beam needs the SPL as injector, but consumes at most $\sim 10\%$
of the SPL protons. The fact that the average neutrino  energies of
both the SuperBeam and the Beta Beam are below 0.5 GeV
(cfr. fig.~\ref{fig:fluxes}), with the Beta
Beam being tunable, offers the fascinating possibility of exposing the
same detector to $2\times 2$ beams (\numu and \nubarmu $\times$
\nue \  and \nubare) having access to
CP, T and CPT searches in the same run.

It is evident that the combination of the two beams would not result    
only in an increase in the statistics of the experiment, 
but it would also offer    
clear advantages in the reduction of the systematic errors, and it would offer
the necessary redundancy to firmly establish any effect of violation of CP within the reach of the experiment.

The CP violation sensitivities of the combined BetaBeam and SPL-SB 
experiments are shown in Fig.~\ref{fig:beta:sens} and  Fig.~\ref{fig:CP:delta}.
\section{Conclusions}
Betabeams are a novel concept which can give very precise insight on the
problem of neutrino mixing. Their design has many common features with the
EURISOL project aiming at producing high intensity radioactive beams  
for nuclear physics studies with astrophysical applications. 
This synergy has been outlined at the Moriond workshop, and common technical
studies are already going on. Recently,
the potentialities of low energy betabeams ($\gamma$ factors below 10)
has also been emphasized \cite{Volpe}.

On the other hand, there have been for some time projects of megaton detectors
to study proton decay and detect supernova explosions, but they never got
financial support. The fact that they would also be 
perfect targets for low
energy neutrino superbeams or betabeams considerably increases the physical
interest of these detectors. Fortunately enough, a possible site exists near
CERN at the right distance and could house as soon as 2015 such a detector.
CERN has a long lasting expertise on ion production and acceleration, and
has announced officially to be site candidate for the Eurisol project.

This offers Europe and CERN a unique opportunity to contribute significantly
to megaton physics, neutrino physics, and Eurisol physics by joining efforts 
of several communities on an ambitious and multidisciplinary project.

\end{document}